\documentclass[preprint, 12pt,authoryear]{elsarticle}

\usepackage{amssymb}
\usepackage{amsthm}
\usepackage{amsmath}

\usepackage{lineno}

\journal{a future journal}

\begin{document}

\begin{frontmatter}


\author{Tsung Fei Khang\corref{cor1}}
\ead{tfkhang@um.edu.my}

\title{A gamma approximation to the Bayesian posterior distribution of a discrete parameter of the Generalized Poisson model}



\address{Institute of Mathematical Sciences, Faculty of Science, University of Malaya, 50603 Kuala Lumpur, Malaysia.}

\begin{abstract}
Let $X$ have a Generalized Poisson distribution with mean $kb$, where $b$ is a known constant in the unit interval and $k$ is a discrete, non-negative parameter. We show that if an uninformative uniform prior for $k$ is assumed, then the posterior distribution of $k$ can be approximated using the gamma distribution when $b$ is small.
\end{abstract}


\end{frontmatter}

\newtheorem{thm}{Theorem}
\newtheorem{cor}{Corollary}
\newtheorem{prop}{Proposition}
\section{Introduction}
\label{introduction}

The family of Generalized Poisson distributions (GP) \citep{consuljain} has been used for more than 40 years to model count data that may be overdispersed or underdispersed. Some of its interesting theoretical properties include a Poisson mixture interpretation \citep{joezhu}, and a heavier tail compared to the negative binomial distribution \citep{joezhu, karlis}. Various chance mechanisms have been found to generate the GP distribution \citep{shoukri}. Numerous applications are given in \cite{consul}, and in recent years, it has gained increasing popularity in bioinformatics for modelling RNA-Seq count data \citep{srivastava, lijiang, wemiq, wang}.

In this work, we study the posterior distribution of a discrete parameter of the GP model under a particular parametrization. Let $X$ be a random variable following $\mbox{GP}(\lambda_1, \lambda_2)$. Its probability mass function (pmf) is given by
\begin{equation}
P(X=x|\lambda_1,\lambda_2) = \frac{\lambda_1(\lambda_1+x\lambda_2)^{x-1} e^{-(\lambda_1+x\lambda_2)}} {x!} \ ,
\label{eq:genpois}
\end{equation}
where $x=0,1,2,\ldots$, $\lambda_1 > 0$ and $|\lambda_2| < 1$. Negative values of $\lambda_2$ correspond to overdispersion, positive values to underdispersion, and $\lambda_2 = 0$ reduces eq.(\ref{eq:genpois}) to the Poisson distribution with mean $\lambda_1$. Consider the following parametrization: $\lambda_2 = 1- \sqrt{m}$, $\lambda_1 = kb\sqrt{m}$, with $m = e^{ab+c}$. We assume that $a,c \in \mathbb{R}$ and $0 < b < 1$ are known constants, and focus our interest on the discrete non-negative parameter $k$. Under this parametrization,
The mean and the variance of the GP model are given by
\begin{eqnarray*}
\mathbb{E}(X|k) &=& \lambda_1 / (1-\lambda_2) = kb\ ,\\
\mbox{Var}(X|k) &=& \mathbb{E}(X|k) / (1-\lambda_2)^2 = kb/m \ , 
\end{eqnarray*}

We are concerned with the posterior probability of the $k$ parameter. In the absence of any prior information, a Bayesian formulation using an improper uniform prior $P(k)=1, k=0,1,2,\ldots$ on $k$ yields the posterior distribution of $k$ as
\begin{equation}
\begin{split}
P(k|X=x) &= \frac{P(X=x|k)}{\sum_{j=x}^\infty P(X=x|j)}\\ 
	&= \frac{k(bk\sqrt{m}+x(1-\sqrt{m}))^{x-1}e^{-bk\sqrt{m}}} {\sum_{j=x}^\infty j(bj\sqrt{m}+x(1-\sqrt{m}))^{x-1}e^{-bj\sqrt{m}}} \\
	&= \frac{k(k+g(x))^{x-1}e^{-bk\sqrt{m}}}{\sum_{j=x}^\infty j(j+g(x))^{x-1}e^{-b\sqrt{m}j}}, \\
	\label{eq:posterior_gen}
	\end{split}
\end{equation}
for $k \geq x$, where $g(x) = \{(1-\sqrt{m}) /(b\sqrt{m})\}x$. It is easy to check that $P(k|X=x)$ is proper even though an improper uniform prior distribution is used. 
The aim of this paper is to derive a continuous approximation to the posterior distribution eq.(\ref{eq:posterior_gen}), so that the posterior mean and variance can be determined directly from the theoretical properties of the approximating distribution.
\section{Results}
\label{results}
\noindent We show that under certain conditions, the gamma distribution approximates the posterior distribution of $k$.
\vskip 5mm
\begin{thm}
\noindent If $P(k|X=x)$ is treated as a density function, then $P(k|X=x)$ is approximately equal to the probability density function of the gamma distribution with mean $(x+1)/(b\sqrt{m})$ and variance $(x+1)/(b^2m)$ for some $k \geq l$ where $l > x$. 
\end{thm}
\vskip 5mm
\begin{proof}\noindent First, we note that the denominator in eq.(\ref{eq:posterior_gen}) can be written as
$$\sum_{j=x}^\infty (j+g(x))^x e^{-b\sqrt{m}j} - g(x)\sum_{j=x}^\infty (j+g(x))^{x-1}e^{-b\sqrt{m}j}.$$
The Lerch transcendent $\Phi(z,s,a)$ is given by
$$\Phi(z,s,a) = \sum_{k=0}^\infty \frac{z^k}{(a+k)^s}, $$ 
where $|z| < 1$, $a \neq 0, -1, -2, \ldots$, and $s \neq 1, 2, \ldots$. Representing the denominator using the Lerch transcendent, we get
\begin{equation}
e^{-b\sqrt{m}x}\Phi(e^{-b\sqrt{m}},-x,wx) - (w-1)xe^{-b\sqrt{m}x}\Phi(e^{-b\sqrt{m}},-(x-1),wx), 
\label{eq:lerchphi}
\end{equation}
where $w = 1+(1-\sqrt{m})/(b\sqrt{m})$.
\vskip 2mm
\noindent The following identity (eq.1.11(11) in \cite{erdelyi}) relates the Lerch transcendent with negative argument for $s$ to the Bernoulli polynomials:
\begin{equation}
\Phi(z,-h,v) = \frac{h!}{z^v}\left( \log\frac{1}{z} \right)^{-(h+1)} - \frac{1}{z^v}\sum_{r=0}^\infty \frac{B_{h+r+1}(v)(\log z)^r}{r!(h+r+1)},
\label{eq:lerchphi_bernoulli} 
\end{equation}
where $|\log z| < 2\pi$, $v \neq 0,-1,-2,\ldots$, $h \neq -1,-2,\ldots$, and $B_n(x)$ is the $n$-th Bernoulli polynomial with argument $x$. The Bernoulli polynomial is defined as
$$B_n(x) = \sum_{j=0}^n {n \choose j} b_{n-j}x^j, $$
where $b_{i}$ is the $i$th Bernoulli number.
\vskip 2mm
\noindent If we now substitute $z=e^{-b\sqrt{m}}$, $h=x$, $v=wx$ into the identity  eq.(\ref{eq:lerchphi_bernoulli}), then we obtain
\begin{eqnarray*}
\Phi(e^{-b\sqrt{m}},-x,wx) &=& \frac{\Gamma(x+1)}{e^{-b\sqrt{m}(wx)}}(b\sqrt{m})^{-(x+1)} \\
&& - \frac{1}{e^{-b\sqrt{m}(wx)}}\sum_{r=0}^\infty \frac{B_{x+r+1}(wx)(-b\sqrt{m})^r}{r!(x+r+1)}
\end{eqnarray*}
For small values of $b$ ($0 < b < 1$), we note that the sum of Bernoulli polynomials is dominated by the zero-th term. Hence, for the first term in eq.(\ref{eq:lerchphi}),
\begin{equation}
e^{-bw\sqrt{m}x}\Phi(e^{-b\sqrt{m}},-x,wx) = \frac{\Gamma(x+1)}{(b\sqrt{m})^{x+1}} - \frac{B_{x+1}(wx)}{x+1} + O(b\sqrt{m}) ,
\label{eq:gamma_bernoulli}
\end{equation}
as $b \rightarrow 0$.
An identity involving the Bernoulli polynomials and the sum of $n$th powers (eq. 1.13(10) in \cite{erdelyi}) gives
$$\frac{B_{n+1}(x)-b_{n+1}}{n+1} = \sum_{r=0}^{x-1}r^n. $$
Since $B_{x+1}(x)$ dominates $b_{x+1}$, eq.(\ref{eq:gamma_bernoulli}) becomes
$$e^{-bw\sqrt{m}x}\Phi(e^{-b\sqrt{m}},-x,wx) \approx \frac{\Gamma(x+1)}{(b\sqrt{m})^{x+1}} -  \sum_{r=0}^{[wx]-1}r^x,
$$
where $[wx]$ is the integer part of $wx$. Here,
$$e^{-bw\sqrt{m}x}\Phi(e^{-b\sqrt{m}},-x,wx) \approx \frac{\Gamma(x+1)}{(b\sqrt{m})^{x+1}} ,$$ 
provided that
$$\sum_{r=0}^{[wx]-1}r^x = o\left(\frac{\Gamma(x+1)}{(b\sqrt{m})^{x+1}}\right),$$
as $x \rightarrow \infty$.
Now,
\begin{eqnarray*}
\frac{(b\sqrt{m})^{x+1}\sum_{r=0}^{[wx]-1}r^x}{\Gamma(x+1)} &<& \frac{b\sqrt{m}  \{(bw\sqrt{m}x)^x + (bw\sqrt{m}x)^x + \cdots + (bw\sqrt{m}x)^x \}}{x!} \\
&=& \frac{(bw\sqrt{m}x)^{x+1}}{x!}
\end{eqnarray*}
Let the upper bound on the right-hand-side be bounded by some constant $\epsilon > 0$, which can be made arbitrarily close to 0. Thus, $(bw\sqrt{m}x)^{x+1}/x! < \epsilon$, and multiplying both sides by $x!$ and then taking logarithm, we get
\begin{equation}
[(1-\sqrt{m} + b\sqrt{m})x] < \exp \left\{ \frac{1}{x+1}\left(\sum_{i=1}^x \log i  + \log \epsilon\right) \right\} .
\label{eq:inequal}
\end{equation}
This implies that for some fixed $\epsilon$, the approximation should be reasonably good as long as $a,b,c$ and $x$ are such that the inequality (\ref{eq:inequal}) is satisfied.

For $a,b,c$ such that $\sqrt{m} \approx 1$, eq.(\ref{eq:lerchphi}) can be approximated as
$$\frac{\Gamma(x+1)}{(b\sqrt{m})^{x+1}} \left(1 - (w-1)x \cdot \frac{b\sqrt{m}}{x} \right) \approx \frac{\Gamma(x+1)}{(b\sqrt{m})^{x+1}}.$$
\noindent It follows that the density function $P(k|X=x)$ (eq.(\ref{eq:posterior_gen})) can be approximated using the probability density function of the gamma distribution with  mean $(x+1)/(b\sqrt{m})$ and variance $(x+1)/(b^2m)$. Thus,
\begin{eqnarray*}
P(k|X=x) &\approx&  \frac{(b\sqrt{m})^{x+1}k(k+g(x))^{x-1}e^{-b\sqrt{m}k}}{\Gamma(x+1)} \\
&=& \frac{k}{k+g(x)} \times \frac{(b\sqrt{m})^{x+1}(k+g(x))^{x+1 - 1}e^{-b\sqrt{m}k}}{\Gamma(x+1)} \\
&\approx& \frac{(b\sqrt{m})^{x+1}k^{x+1-1}e^{-b\sqrt{m}k}}{\Gamma(x+1)},
\end{eqnarray*}
for $k \geq l$, since $g(x) = \{(1-\sqrt{m}) /(b\sqrt{m})\}x$ and there exists some $l > x$, such that $k + g(x) = k + o(k)$ when $k \geq l$.

\end{proof}
\vskip 5mm
\begin{cor}
\noindent For $k \geq l$, where $l > x$, the probability mass function $P(k|X=x)$ can be approximated as
\begin{eqnarray*}
P(k|X=x) &\approx& \int_{k-0.5}^{k+0.5} \frac{(b\sqrt{m})^{x+1}t^{x+1-1}e^{-b\sqrt{m}t}}{\Gamma(x+1)}dt \\
&=& \frac{1}{\Gamma(x+1)}\{ \gamma(x+1,b\sqrt{m}(k+1/2)) - \gamma(x+1,b\sqrt{m}(k-1/2)) \},
\end{eqnarray*}
where $\gamma(u,v)$ is the lower incomplete gamma function:
$$\gamma(u,v) = \int_0^vt^{u-1}e^{-t}dt.$$
\end{cor}
\vskip 5mm
\subsection{Computational validation}
The preceding results establish that the gamma family of distribution is a valid approximation to eq.(\ref{eq:posterior_gen}). Computational validation result suggests that an improvement to the fit can be obtained by replacing the shape and scale parameters of the gamma distribution in Theorem 1 as follows. Let $\mu_{post}$ and $\sigma^2_{post}$ be the posterior mean and the posterior variance, respectively. For a gamma distribution with shape parameter $\alpha$ and scale parameter $\beta$, its mean and variance are given by $\alpha \beta$ and $\alpha \beta^2$, respectively. Given $a,b,c$ and $x$, we can compute the exact mean and variance of the posterior distribution of $k$:
\begin{eqnarray*}
\mu_{post} &=& \sum_{k=x}^\infty k P(k|X=x), \\
\sigma^2_{post} &=& \sum_{k=x}^\infty k^2 P(k|X=x) - \mu^2_{post}.
\end{eqnarray*}
Simple algebra then yields $\alpha = \mu^2_{post} / \sigma^2_{post}$ and $\beta = \sigma^2_{post}/\mu_{post}$. The gamma approximation with $\alpha$ and $\beta$ thus computed fits eq.(\ref{eq:posterior_gen}) better than $\alpha = x+1$, $\beta = 1/(b\sqrt{m})$ given in Theorem 1. Figure 1 shows the how well the gamma approximation fits the posterior distribution of $k$ with and without adjustment to $\alpha$ and $\beta$ parameters. In both cases the quality of the gamma approximation deteriorates when $x$ becomes relatively large if the $\alpha$ and $\beta$ parameters are not adjusted.
\begin{figure}
\vskip -5cm
\centering
{
\includegraphics[scale=0.65]{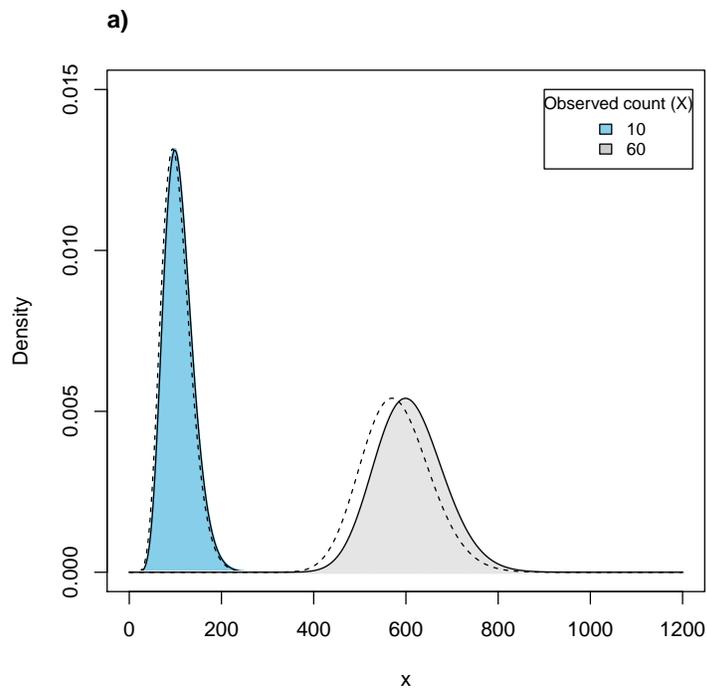}
\includegraphics[scale=0.65]{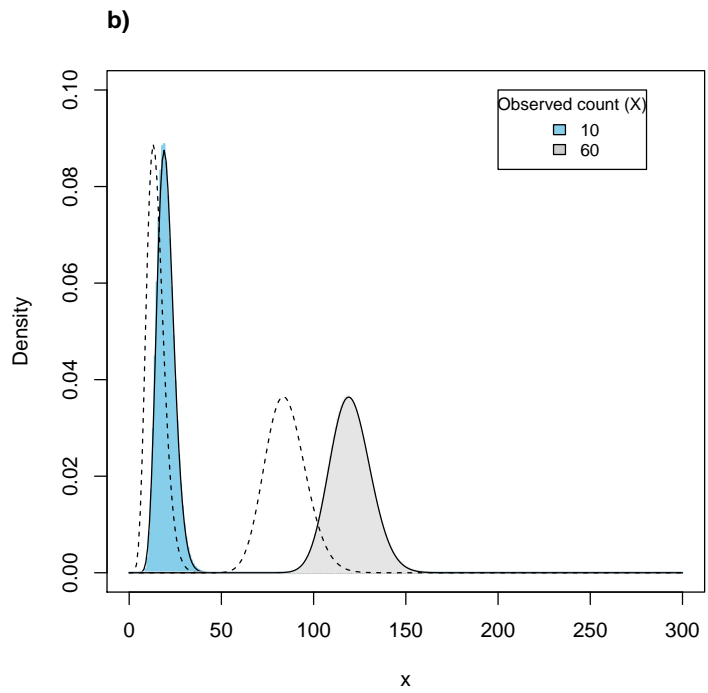}
}
\caption{Examples of fitting the gamma distribution to the posterior distribution of $k$ for two cases: a) relatively small $b$ parameter: $a=1.5, b=0.1, c=-0.05$ ($m = 1.1$); b) relatively large $b$ parameters: $a=1.5, b=0.5, c=-0.05$ ($m=2.0$). Solid lines indicate gamma approximation with shape and scale parameters determined by matching moments of the posterior distribution of $k$ and the gamma distribution. Broken lines indicate gamma approximation with shape and scale parameters given in Theorem 1. }
\label{fig:gammafit}
\end{figure}

\section{Acknowledgment}
\label{acknowledgment}
I thank Martti Tammi and Joel Zi-Bin Low for stimulating the current work by discussing an applied problem in RNA-Seq count data analysis with me. Vanamamalai Seshadri read and provided feedback on the initial draft.


\pagebreak

\bibliographystyle{elsarticle-num-names} 
\bibliography{Gamma_approximation_GP_KTF}


\end{document}